\documentclass[dvips,12pt,a4paper]{article}
\usepackage{a4wide}
\usepackage{epsfig}
\usepackage{amsmath}
\usepackage{latexsym}
\usepackage{array}

  \newlength{\absize}
\newcommand{\dd}{\mbox{{\rm d}}}

\def\lsim{\mathrel{\rlap{\raise 2.5pt \hbox{$<$}}\lower 2.5pt
\hbox{$\sim$}}}

\newcommand{\Lumint}{{\cal L}_{\rm int}}

\allowdisplaybreaks 

\catcode`@=11
\def\citer{\@ifnextchar [{\@tempswatrue\@citexr}{\@tempswafalse\@citexr[]}}

\def\@citexr[#1]#2{\if@filesw\immediate\write\@auxout{\string\citation{#2}}\fi
  \def\@citea{}\@cite{\@for\@citeb:=#2\do
    {\@citea\def\@citea{--\penalty\@m}\@ifundefined
       {b@\@citeb}{{\bf ?}\@warning
       {Citation `\@citeb' on page \thepage \space undefined}}%
\hbox{\csname b@\@citeb\endcsname}}}{#1}}
\catcode`@=12

\begin{document}
  \thispagestyle{empty}
  \pagestyle{empty}
  \renewcommand{\thefootnote}{\fnsymbol{footnote}}
\newpage\normalsize
    \pagestyle{plain}
    \setlength{\baselineskip}{4ex}\par
    \setcounter{footnote}{0}
    \renewcommand{\thefootnote}{\arabic{footnote}}
\newcommand{\preprint}[1]{%
  \begin{flushright}
    \setlength{\baselineskip}{3ex} #1
  \end{flushright}}
\renewcommand{\title}[1]{%
  \begin{center}
    \LARGE #1
  \end{center}\par}
\renewcommand{\author}[1]{%
  \vspace{2ex}
  {\Large
   \begin{center}
     \setlength{\baselineskip}{3ex} #1 \par
   \end{center}}}
\renewcommand{\thanks}[1]{\footnote{#1}}
\renewcommand{\abstract}[1]{%
  \vspace{2ex}
  \normalsize
  \begin{center}
    \centerline{\bf Abstract}\par
    \vspace{2ex}
    \parbox{\absize}{#1\setlength{\baselineskip}{2.5ex}\par}
  \end{center}}

\begin{flushright}
{\setlength{\baselineskip}{2ex}\par

}
\end{flushright}
\vspace*{4mm}
\vfill
\title{
Discriminating graviton exchange effects from other new physics 
scenarios in $e^+e^-$ collisions
}
\vfill
\author{
P. Osland$^{a}$, A.A. Pankov$^{a,b,c}$ {\rm and} N. Paver$^{c}$}
\begin{center}
$^a$ Department of Physics, University of Bergen, 
N-5007 Bergen, Norway\\ 
$^b$ Pavel Sukhoi Technical University, 
     Gomel 246746, Belarus \\
$^c$ University of Trieste  and INFN-Sezione di Trieste, 34100 
Trieste, Italy \\

\end{center}
\vfill 
\abstract 
{We study the possibility of uniquely identifying the effects of 
graviton exchange from other new physics in high energy 
$e^+e^-$ annihilation into fermion-pairs. For this purpose, we use as 
basic observable a specific asymmetry among integrated differential 
distributions, that seems particularly suitable to directly test for such 
gravitational effects in the data analysis.}

\vspace*{20mm}
\setcounter{footnote}{0}
\vfill

\newpage
    \setcounter{footnote}{0}
    \renewcommand{\thefootnote}{\arabic{footnote}}

\section{Introduction}
All types of new physics (NP) scenarios are determined by non-standard
dynamics involving new building blocks and forces mediated by exchange
corresponding to heavy states with mass scales $\Lambda$ much greater than
$M_{W}$. Unambiguous confirmation of such dynamics would require the
experimental discovery of the envisaged new heavy objects and the measurement
of their coupling constants to ordinary quarks and leptons. While there is
substantial belief that the supersymmetric partners of the Standard Model (SM)
particles should be directly produced, and identified, at future
proton--proton and electron--positron high energy colliders such as the LHC
and the Linear Collider (LC), in the other cases the current experimental
limits on the new, heavy particles are so high, of the order of several (or
tens of) TeV, that one cannot expect them to be directly produced at the
energies foreseen for these machines. In this situation, the new interactions
can manifest themselves only by indirect, virtual, effects represented by
deviations of the measured observables from the SM numerical predictions. The
problem, then, is to identify from the data analysis the possible new
interactions, because different NP scenarios can in principle cause similar
measurable deviations, and for this purpose suitable observables must be
defined.  \par

At ``low'' energies (compared to the above-mentioned large mass scales) the
physical effects of the new interactions are conveniently accounted for, in
reactions involving the familiar quarks and leptons, by effective {\it
contact-interaction} (CI) Lagrangians that provide the expansion of the
relevant transition amplitudes to leading order in the small ratio $\sqrt
s/\Lambda$ ($\sqrt s$ being the c.m. energy).  
\par

Familiar classes of contact interactions are represented by composite models
of quarks and leptons \cite{'tHooft:xb,Eichten:1983hw}; exchanges of very
heavy $Z^\prime$ with a few TeV mass \cite{Barger:1997nf,Hewett:1988xc} and of
scalar and vector heavy leptoquarks \cite{Buchmuller:1986zs}; in the SUSY
context, $R$-parity breaking interactions mediated by sneutrino exchange
\cite{Kalinowski:1997bc,Rizzo:1998vf}; bi-lepton boson exchanges
\cite{Cuypers:1996ia}; anomalous gauge boson couplings (AGC)
\cite{Gounaris:1997ft}; virtual Kaluza--Klein (KK) graviton exchange in the
context of gravity propagating in large extra dimensions, exchange of gauge
boson KK towers or string excitations, {\it etc.}
\citer{Arkani-Hamed:1998rs,Cheung:2001mq}. Of course, this list is not
exhaustive, because other kinds of {\it contact interactions} may well exist.
\par

In this note, we briefly discuss the deviations induced by {\it contact
interactions} in the electron--positron annihilation into fermion pairs at the
planned Linear Collider energies
\cite{Aguilar-Saavedra:2001rg,Assmann:2000hg}.  In particular, we propose a
simple observable that can be used to unambiguously identify graviton KK tower
exchange effects in the data, relying on its spin-two character and by
``filtering'' out contributions of other NP interactions.  \par

If deviations from the SM predictions were effectively measured, the
identification of the NP source could be attempted by Monte Carlo best fits of
the observed effects, and this would apply also to graviton exchange
\cite{Pasztor:2001hc}. Alternatively, moments of the differential cross
section folded with Legendre polynomial weights appear to be a promising
technique to pin down NP effects in the case of electron--positron reactions
induced at the SM level by $s$-channel exchanges \cite{Rizzo:2002pc}. Here, we
shall consider a suitably defined combination of integrated cross sections,
the so-called ``center--edge'' asymmetry $A_{\rm CE}$, that allows to
disentangle the graviton exchange in a very simple, and efficient, way.
Specifically, in Sect.~2 we present the required kinematical details and
discuss the properties of $A_{\rm CE}$, in Sect.~3 we discuss beam
polarization, in Sect.~4 we evaluate the sensitivity of this observable to the
characteristic mass parameter of the graviton KK tower exchange, in Sect.~5 we
find the corresponding identification reaches and discuss an application to
sneutrino exchange, differentiating it from KK graviton exchange and, finally,
Sect.~6 is devoted to some comments and conclusive remarks.
               
\section{The center--edge asymmetry \boldmath$A_{\rm CE}$ }

We consider the process (with $f\ne e,t$) 
\begin{equation}
e^++e^-\to f+\bar{f}, \label{proc} \end{equation}   
and, neglecting all fermion masses with respect to $\sqrt s$, we can write 
the differential angular distribution for unpolarized $e^+e^-$ beams in 
terms of $s$-channel $\gamma$ and $Z$ exchanges plus any 
{\it contact-interaction} terms in the following form \cite{Schrempp:1987zy}: 
\begin{equation}
\frac{\dd\sigma}{\dd z}=\frac{1}{4}\left(
\frac{\dd\sigma_{\rm LL}}{\dd z}+
\frac{\dd\sigma_{\rm RR}}{\dd z}+
\frac{\dd\sigma_{\rm LR}}{\dd z}+
\frac{\dd\sigma_{\rm RL}}{\dd z}\right).
\label{crossdif}
\end{equation}
Here, $z\equiv\cos\theta$, with $\theta$ the angle between the 
incoming electron and the outgoing fermion in the c.m.\ frame, and  
$\dd\sigma_{\alpha\beta}/{\dd\cos\theta}$ ($\alpha,\beta={\rm L,R}$) are the 
helicity cross sections given by:
\begin{equation}
\frac{\dd\sigma_{\alpha\beta}}{\dd z}=
N_C\hskip 2pt \frac{\pi\alpha_{\rm e.m.}^2}{2s}\,  
\vert {\cal M}_{\alpha\beta}\vert^2\,
(1\pm z)^2,
\label{helcross}
\end{equation}
where the two signs $\pm$ correspond to the $\rm LL$, $\rm RR$, and $\rm LR$,
$\rm RL$, helicity configurations, respectively, and $N_C\simeq 3
(1+\alpha_s/\pi)$ represents the number of colours of the final state,
including the first-order QCD correction. The helicity amplitudes 
${\cal M}_{\alpha\beta}$ can be written as
\begin{equation}
{\cal M}_{\alpha\beta}={\cal M}_{\alpha\beta}^{\rm SM}+\Delta_{\alpha\beta}
=Q_e Q_f+g_\alpha^e\,g_\beta^f\,\chi_Z+\Delta_{\alpha\beta},
\label{amplit}
\end{equation}
where: $\chi_Z=s/(s-M^2_Z+iM_Z\Gamma_Z)\approx s/(s-M^2_Z)$ represents the $Z$
propagator; $g_{\rm L}^f=(I_{3L}^f-Q_f s_W^2)/s_W c_W$ and $g_{\rm R}^f=-Q_f
s_W/c_W$ are the SM left- and right-handed fermion couplings of the $Z$
with $s_W^2=1-c_W^2\equiv \sin^2\theta_W$; $Q_e$ and $Q_f$ are the fermion
electric charges.  The $\Delta_{\alpha\beta}$ functions represent the contact
interaction contributions coming from TeV-scale physics. 

The structure of the differential cross section
(\ref{crossdif})--(\ref{amplit}) is particularly interesting in that it is
equally valid for a wide variety of New Physics (NP) models listed in
Table~\ref{table:epsilon}. Note that only graviton exchange induces a modified
angular dependence to the differential cross section via its $z$-dependence of
$\Delta_{\alpha\beta}$.

We define the generalized center--edge asymmetry $A_{\rm CE}$ as 
\cite{Pankov:1997da}:
\begin{equation}
A_{\rm CE}=\frac{\sigma_{\rm CE}}{\sigma},
\label{ace} 
\end{equation} 
in terms of the difference between the central and edge parts
of the cross section
\begin{equation}
\sigma_{\rm CE}=\left[\int_{-z^*}^{z^*}-
\left(\int_{-1}^{-z^*}+\int_{z^*}^{1}\right)\right]
\frac{\dd\sigma}{\dd z}\,{\dd z}, 
\label{sce} 
\end{equation} 
and the total cross section
\begin{equation}
\sigma=\int_{-1}^{1}
\frac{\dd\sigma}{\dd z}\,{\dd z},
\label{sigmatot} 
\end{equation}
and $0<z^*<1$.\footnote{The center--edge asymmetry $A_{\rm CE}$ 
for $W$-pair production and fixed $z^*=0.5$ 
has been introduced in \cite{Gounaris:1992kp}.}

\begin{table}[htb]
\centering
\caption{Parametrization of the $\Delta_{\alpha\beta}$ functions in different
models ($\alpha,\beta={\rm L,R}$).}
\vspace*{8pt}
\setlength{\extrarowheight}{6pt}
\begin{tabular}{|c|c|}
\hline
Model & $\Delta_{\alpha\beta}$  \\  \hline\hline
composite fermions \cite{Eichten:1983hw} & 
${\displaystyle{\pm\frac{s}{\alpha_{\rm e.m.}}\frac{1}
{\Lambda_{\alpha\beta}^2}}}$ \\  \hline
extra gauge boson $Z'$ \cite{Barger:1997nf,Hewett:1988xc} & 
${\displaystyle{g_\alpha'{}^e\,g_\beta'{}^f\,\chi_{Z'}}}$ \\ \hline 
AGC ($f=\ell$) \cite{Gounaris:1997ft} & 
${\displaystyle{\Delta_{\rm LL}=s\left(\frac{\tilde f_{DW}}{2s_W^2}+
\frac{2 \tilde f_{DB}}{c_W^2}\right)}}$,
${\displaystyle{\frac{\Delta_{\rm RR}}{2}=\Delta_{\rm LR}=\Delta_{\rm RL}
=s\frac{4 \tilde f_{DB}}{c_W^2}}}$  \\  \hline
TeV-scale extra dim. \cite{Pasztor:2001hc,Cheung:2001mq} & 
${\displaystyle{(Q_eQ_f+g_\alpha^e\,g_\beta^f)\,\frac{\pi^2}{3\,M_C^2}}}$ 
\\  \hline
ADD model \cite{Arkani-Hamed:1998rs,Hewett:1998sn} & 
${\displaystyle{\Delta_{\rm LL}=\Delta_{\rm RR}=f_G\,(1-2\,z)}}$,  
${\displaystyle{\Delta_{\rm LR}=\Delta_{\rm RL}=-f_G\,(1+2\,z)}}$
\\  \hline
\end{tabular}
\label{table:epsilon}
\end{table}
In Table~\ref{table:epsilon} $\Lambda_{\alpha\beta}$ are compositeness scales;
$\chi_{Z'}$ is the $Z'$ propagator defined according to $\chi_Z$; $\tilde
f_{DW}$ and $\tilde f_{DB}$ are related to $f_{DW}$ and $f_{DB}$ of
ref.~\cite{Gounaris:1997ft} by $\tilde f=f/m_t^2$ ($f_{DW}$ and $f_{DB}$
parametrize new-physics effects associated with the SU(2) and hypercharge
currents, respectively); $M_C$ is the compactification scale; finally,
$f_G=\lambda\,s^2/(4\pi\alpha_{\rm e.m.}M_H^4)$ parametrizes the strength
associated with massive graviton exchange with $M_H$ the cut-off scale in the
KK graviton tower sum.  Note that, compared with, e.g., the composite fermion
case, the KK graviton effect is suppressed by the (larger) power $(\sqrt
s/M_H)^4$, so that a lower reach on $M_H$ can be expected in comparison to the
constraints obtainable, at the same c.m.\ energy, on $\Lambda$'s. The effect
of the extra dimensional model \cite{Pasztor:2001hc} is $s$-independent, and
the sign of $\Delta_{\alpha\beta}$ is fixed.

First, let us consider graviton exchange effects.  For definiteness we
consider the ADD model \cite{Arkani-Hamed:1998rs}.  From
Eqs.~(\ref{crossdif})--(\ref{sigmatot}) and Table~\ref{table:epsilon} one
can derive the asymmetry $A_{\rm CE}$ for the process (\ref{proc}) including
graviton tower exchange:
\begin{equation}
A_{\rm CE}=\frac{\sigma_{\rm CE}^{\rm SM}+ \sigma_{\rm CE}^{\rm INT}
+\sigma_{\rm CE}^{\rm NP}}
{\sigma^{\rm SM}+\sigma^{\rm INT}+\sigma^{\rm NP}},
\label{aceG} 
\end{equation}   
where ``SM'', ``INT'' and ``NP'' refer to ``Standard Model'', ``Interference''
and (pure) ``New Physics'' contributions.  Explicitly, we have
\begin{align}   \label{Eq:sigmas}
\sigma_{\rm CE}^{\rm SM}
=&N_C\hskip 2pt \frac{\pi\alpha_{\rm e.m.}^2}{2s}\,  
\frac{1}{4}\left[({\cal M}_{\rm LL}^{\rm SM})^2+({\cal M}_{\rm RR}^{\rm SM})^2
+({\cal M}_{\rm LR}^{\rm SM})^2+({\cal M}_{\rm RL}^{\rm SM})^2\right]\,
\frac{4}{3}\left[z^*({z^*}^2+3)-2\right],  \nonumber \\
\sigma_{\rm CE}^{\rm INT}
=&N_C\hskip 2pt \frac{\pi\alpha_{\rm e.m.}^2}{2s}\,2\,f_G\,  
\frac{1}{4}\left[{{\cal M}_{\rm LL}^{\rm SM}}+{{\cal M}_{\rm RR}^{\rm SM}}
-{{\cal M}_{\rm LR}^{\rm SM}}-{{\cal M}_{\rm RL}^{\rm SM}}\right]\,
4z^*(1-{z^*}^2),  \nonumber \\
\sigma_{\rm CE}^{\rm NP}
=&N_C\hskip 2pt \frac{\pi\alpha_{\rm e.m.}^2}{2s}\,f_G^2\,  
\frac{4}{5}    
\left[4{z^*}^5+5z^*(1-{z^*}^2)-2\right],
\end{align}
with
\begin{align}
\sigma^{\rm SM}
=&N_C\hskip 2pt \frac{\pi\alpha_{\rm e.m.}^2}{2s}\,  
\frac{1}{4}\left[({\cal M}_{\rm LL}^{\rm SM})^2+({\cal M}_{\rm RR}^{\rm SM})^2
+({\cal M}_{\rm LR}^{\rm SM})^2+({\cal M}_{\rm RL}^{\rm SM})^2\right]\,
\frac{8}{3},   \nonumber \\
\sigma^{\rm INT}=&0, \qquad
\sigma^{\rm NP}
=N_C\hskip 2pt \frac{\pi\alpha_{\rm e.m.}^2}{2s}\,f_G^2\,  
\frac{8}{5}.
\label{sNP} 
\end{align}   
Note that, at $z^*=0$ and 1, $\sigma_{\rm CE}=\mp\sigma$, respectively.

In the case of the SM the center--edge asymmetry $A_{\rm CE}^{\rm SM}$ can be
obtained from Eqs.~(\ref{aceG})--(\ref{sNP}) taking $f_G=0$:
\begin{equation}
A_{\rm CE}^{\rm SM}=\frac{\sigma_{\rm CE}^{\rm SM}}
{\sigma^{\rm SM}}=\frac{1}{2}\,z^*\,({z^*}^2+3)-1.
\label{aceSM}
\end{equation}
It is interesting to note that in Eq.~(\ref{aceSM}) the helicity amplitudes in
the numerator and denominator cancel and only a ratio of kinematical factors
remains in the limit of neglecting external fermion masses. In addition,
$A_{\rm CE}^{\rm SM}$ is independent of energy and of the flavour of the
final-state fermions.  It contains {\it only} the kinematical variable $z^*$.
Fig.~\ref{Fig:fig1} shows $A_{\rm CE}^{\rm SM}$ as a function of $z^*$.  From
Eq.~(\ref{aceSM}) one can determine the value of $z^*$ where $A_{\rm CE}^{\rm
SM}$ vanishes \cite{Datta:2002tk},
\begin{equation}
z^*_0=(\sqrt{2}+1)^{1/3}-(\sqrt{2}-1)^{1/3}=0.596,
\label{z*0}
\end{equation}
corresponding to $\theta=53.4^\circ$ (see the solid curve in
Fig.~\ref{Fig:fig1}).

\begin{figure}[htb]
\refstepcounter{figure}
\label{Fig:fig1}
\addtocounter{figure}{-1}
\begin{center}
\setlength{\unitlength}{1cm}
\begin{picture}(10.0,6.0)
\put(0,-0.5)
{\mbox{\epsfysize=7.5cm
\epsffile{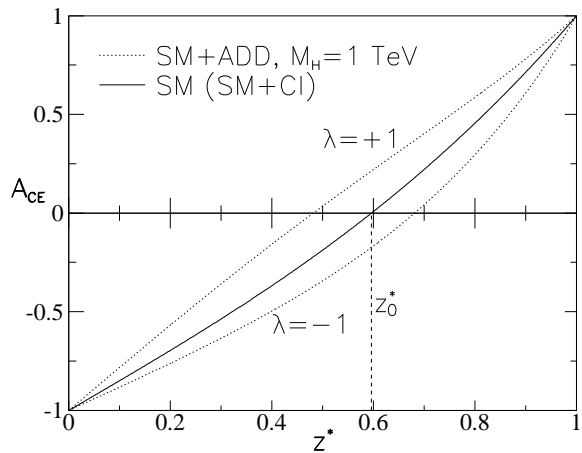}}}
\end{picture}
\caption{Tree-diagram result for $A_{CE}$ for the process $e^+e^-\to l^+l^-$
($l=\mu,\tau$) as a function of $z^*$ in the SM and in the ADD model with
$M_H=1$ TeV and $\lambda=\pm 1$.}
\end{center}
\end{figure}

Graviton exchange in the ADD model affects  $A_{CE}$ inducing a deviation 
from the SM prediction:
\begin{equation}
\Delta A_{\rm CE}=A_{\rm CE}-A_{\rm CE}^{\rm SM}.
\label{deviat} 
\end{equation} 
For $(s/M_H^2)^2\ll1$, it will be $\sigma_{\rm CE}^{\rm INT}$ which will
produce the largest deviation from the expectations of the SM, since this term
is of order $(\sqrt s/M_H)^4$, whereas the pure NP contribution proportional
to $f_G^2$ in Eqs.~(\ref{Eq:sigmas}) and (\ref{sNP}) is of the much higher
order $(\sqrt s/M_H)^8$.  Taking into account only SM-NP interference terms,
one derives:
\begin{equation}  
\Delta A_{\rm CE}\simeq f_G\,\frac{  
 {\cal M}_{\rm LL}^{\rm SM}+{\cal M}_{\rm RR}^{\rm SM}
-{\cal M}_{\rm LR}^{\rm SM}-{\cal M}_{\rm RL}^{\rm SM}}
{\left[({\cal M}_{\rm LL}^{\rm SM})^2+({\cal M}_{\rm RR}^{\rm SM})^2
+({\cal M}_{\rm LR}^{\rm SM})^2+({\cal M}_{\rm RL}^{\rm SM})^2\right]}\,
3\,z^*\,({1-z^*}^2).
\label{delace}
\end{equation}

For the lepton pair production process in the ADD model, the corresponding
$A_{\rm CE}$ is shown in Fig.~\ref{Fig:fig1} for $M_H=1$ TeV and $\lambda=\pm
1$. As one can see from Fig.~\ref{Fig:fig1}, $\Delta A_{\rm CE}=0$ for $z^*=0$
and 1. Clearly, in contrast to $A_{\rm CE}^{\rm SM}$, the $\Delta A_{\rm CE}$
of Eq.~(\ref{delace}) depends on the flavour of the final-state fermion $f$.

To illustrate the effect of graviton exchange on the center--edge asymmetry,
we show in Fig.~\ref{Fig:fig2} the $z^*$-distributions of the deviation
$\Delta A_{\rm CE}$, taking as examples the values of $M_H$ indicated in the
caption.  The deviation $\Delta A_{\rm CE}$ [including also the pure NP term
in addition to the simple result of Eq.~(\ref{delace})] is compared to the
expected statistical uncertainties, $\delta A_{\rm CE}$, represented by the
vertical bars and given by
\begin{equation}
\delta A_{\rm CE}=\sqrt{\frac{1- (A_{\rm CE}^{\rm SM})^2}
{\epsilon_f\,\Lumint\, \sigma^{\rm SM}}}.
\label{uncert}
\end{equation}
Here, $\Lumint$ is the integrated luminosity, and $\epsilon_f$ the efficiency
for reconstruction of $f\bar{f}$ pairs.  We will assume that the efficiencies
of identifying the final state fermions are rather high: 100\% for
$l=\mu,\,\tau$, 80\% for $f=b$, and 60\% for $f=c$.  Fig.~\ref{Fig:fig2}
qualitatively indicates that, for the chosen values of the c.m. energy $\sqrt
s$ and $\Lumint$, the reach on $M_H$ will be of the order of 2.5 TeV.
\begin{figure}[htb]
\refstepcounter{figure}
\label{Fig:fig2}
\addtocounter{figure}{-1}
\begin{center}
\setlength{\unitlength}{1cm}
\begin{picture}(10.0,6.0)
\put(0,-0.5)
{\mbox{\epsfysize=7.5cm
\epsffile{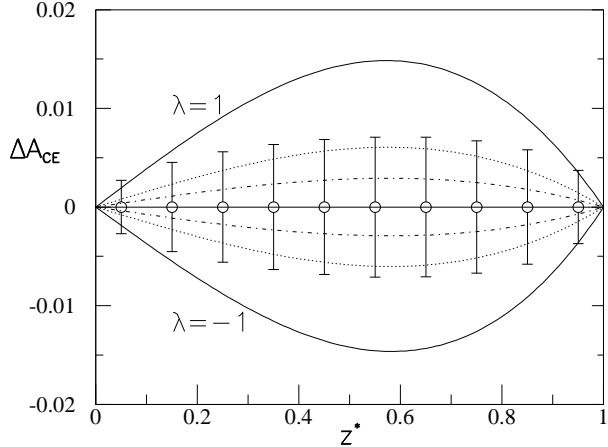}}}
\end{picture}
\caption{The deviation of $A_{\rm CE}$ [cf.\ Eq.~(\ref{deviat})] from the SM
(or SM+CI) expectations (at tree level) as a function of $z^*$ for the process
$e^+e^-\to l^+l^-$ for $M_H=2$ (solid), 2.5 (dotted), and 3 TeV (dash-dotted),
$\lambda=\pm 1$ and $\sqrt s=0.5$ TeV.  The expected statistical uncertainties
at $\Lumint=50\ \mbox{fb}^{-1}$ are shown as error bars.}
\end{center}
\end{figure}

Now, let us consider the conventional {\it contact-interaction}-like effects
parametrized by $z$-independent $\Delta_{\alpha\beta}$ summarized in Table~1.
Application of Eq.~(\ref{ace}) to composite-like contact interactions is
straightforward, the result can be written as:
\begin{equation}
A_{\rm CE}^{\rm SM+CI}=\frac{\sigma_{\rm CE}^{\rm SM+CI}}
{\sigma^{\rm SM+CI}},
\label{aceCI} 
\end{equation}   
where
\begin{equation}
\sigma_{\rm CE}^{\rm SM+CI}
=N_C\hskip 2pt \frac{\pi\alpha_{\rm e.m.}^2}{2s}\,  
\frac{1}{4}\left[({\cal M}_{\rm LL})^2+({\cal M}_{\rm RR})^2
+({\cal M}_{\rm LR})^2+({\cal M}_{\rm RL})^2\right]\,
\frac{4}{3}\left[z^*({z^*}^2+3)-2\right],
\label{sceCI} 
\end{equation}   
and
\begin{equation}
\sigma^{\rm SM+CI}
=N_C\hskip 2pt \frac{\pi\alpha_{\rm e.m.}^2}{2s}\,  
\frac{1}{4}\left[({\cal M}_{\rm LL})^2+({\cal M}_{\rm RR})^2
+({\cal M}_{\rm LR})^2+({\cal M}_{\rm RL})^2\right]\,
\frac{8}{3}.
\label{sCI} 
\end{equation}   
From Eqs.~(\ref{aceCI})--(\ref{sCI}), one has
\begin{equation}
A_{\rm CE}^{\rm SM+CI}=\frac{1}{2}\,z^*\,({z^*}^2+3)-1.
\label{ACECI}
\end{equation}   

This result is {\it identical} to $A_{\rm CE}^{\rm SM}$ defined by
Eq.~(\ref{aceSM})!  In other words, $A_{\rm CE}$ has the form (\ref{ACECI}) in
the SM and will remain so even if {\it contact-interaction}-like effects are
present.  Thus, conventional contact-interaction effects, being described by
current--current interactions, yield {\it the same} center--edge asymmetry as
the Standard Model. The reason is simply that both these interactions are
described by vector currents, as opposed to the tensor couplings of gravity.
The deviation of $A_{\rm CE}$ from the SM (and SM+CI) prediction is clearly a
signal of the spin-2 particle exchange. Thus, it is clear that a non-zero
value of $\Delta A_{\rm CE}$ can provide a clean signature for graviton, or
more generally, spin-2 exchange in the process $e^+e^-\to\bar{f}f$.

\section{Polarized beams}
Let us now consider the case of longitudinally polarized beams, with $P$ and
$\bar P$ the degrees of polarization of the electron and positron beams,
respectively. The polarized differential cross section can then be written as
\begin{equation}
\frac{\dd\sigma}{\dd z}=\frac{D}{4}
\left[(1-P_{\rm eff})
\left(\frac{\dd\sigma_{\rm LL}}{\dd z}
+\frac{\dd\sigma_{\rm LR}}{\dd z}\right)
+(1+P_{\rm eff})\left(\frac{\dd\sigma_{\rm RR}}{\dd z}
+\frac{\dd\sigma_{\rm RL}}{\dd z}\right)\right],
\label{crossdif-pol}
\end{equation}
where $D=1-P\bar{P}$ and $P_{\rm eff}=(P-\bar{P})/(1-P\bar{P})$ is the
effective polarization \cite{Flottmann:1995ga}.  For example, $P_{\rm eff}=\pm
0.95$ and $D\approx 1.5$ for $P=\pm 0.8$ and $\bar{P}=\mp 0.6$.

In addition, in the case of a reduced kinematical region, with cuts around the
beam pipe, $\vert z\vert\le z_{\rm cut}$ ($0<z_{\rm cut}<1$), one can define
the generalized center--edge asymmetry $A_{\rm CE}$ as above, with
Eqs.~(\ref{sce}) and (\ref{sigmatot}) replaced by
\begin{equation}
\sigma_{\rm CE}=\left[\int_{-z^*}^{z^*}-
\left(\int_{-z_{\rm cut}}^{-z^*}+\int_{z^*}^{z_{\rm cut}}\right)\right]
\frac{\dd\sigma}{\dd z}\,{\dd z}, 
\label{sce-cut} 
\end{equation} 
and 
\begin{equation}
\sigma=\int_{-z_{\rm cut}}^{z_{\rm cut}}
\frac{\dd\sigma}{\dd z}\,{\dd z},
\label{sigmatot-cut} 
\end{equation}
with $0<z^*<z_{\rm cut}$. 

Allowing for angular cuts, as discussed above, the asymmetry $A_{\rm CE}$
including graviton tower exchange can for polarized beams be expressed as
given by Eq.~(\ref{aceG}), with
\begin{align}
\sigma_{\rm CE}^{\rm SM}(z^*,z_{\rm cut})
=&N_C\hskip 2pt \frac{\pi\alpha_{\rm e.m.}^2}{2s}\,  
\frac{D}{4}\left\{(1-P_{\rm eff})\left[({\cal M}_{\rm LL}^{\rm SM})^2
+({\cal M}_{\rm LR}^{\rm SM})^2\right]
+(1+P_{\rm eff})\left[({\cal M}_{\rm RR}^{\rm SM})^2
+({\cal M}_{\rm RL}^{\rm SM})^2\right]\right\}\nonumber \\
&\times F^{\rm SM}(z^*,z_{\rm cut}),  \nonumber \\
\sigma_{\rm CE}^{\rm INT}(z^*,z_{\rm cut})
=&N_C\hskip 2pt \frac{\pi\alpha_{\rm e.m.}^2}{2s}\,2\,f_G\,  
\frac{D}{4}\left[(1-P_{\rm eff})\left({\cal M}_{\rm LL}^{\rm SM}
-{\cal M}_{\rm LR}^{\rm SM}\right)
+(1+P_{\rm eff})
\left({\cal M}_{\rm RR}^{\rm SM}-{\cal M}_{\rm RL}^{\rm SM}\right)\right]
\nonumber \\
&\times F^{\rm INT}(z^*,z_{\rm cut}), \nonumber \\
\sigma_{\rm CE}^{\rm NP}(z^*,z_{\rm cut})
=&N_C\hskip 2pt \frac{\pi\alpha_{\rm e.m.}^2}{2s}\,f_G^2\,D\,  
F^{\rm NP}(z^*,z_{\rm cut}).
\label{scei}
\end{align}
Here, the dependences on the parameter $z^*$ and on the angular cut $z_{\rm
cut}$, are given by
\begin{align}
F^{\rm SM}(z^*,z_{\rm cut})
=&\frac{2}{3}\left[2z^*({z^*}^2+3)-z_{\rm cut}(z_{\rm cut}^2+3)\right],
\nonumber \\
F^{\rm INT}(z^*,z_{\rm cut})
=&2\left[2z^*(1-{z^*}^2)-z_{\rm cut}(1-z_{\rm cut}^2)\right],
\nonumber \\
F^{\rm NP}(z^*,z_{\rm cut})
=&\frac{2}{5}\left[8{z^*}^5+10z^*(1-{z^*}^2)-4z_{\rm cut}^5
-5z_{\rm cut}(1-z_{\rm cut}^2)\right].
\label{F} 
\end{align}
The total cross sections in the denominator of Eq.~(\ref{aceG}) 
can be derived from Eqs.~(\ref{scei}) and (\ref{F}):
\begin{equation}
\sigma^{\rm SM}(z_{\rm cut})=\sigma_{\rm CE}^{\rm SM}(z^*=z_{\rm cut}),\quad
\sigma^{\rm INT}(z_{\rm cut})=\sigma_{\rm CE}^{\rm INT}(z^*=z_{\rm cut}),\quad
\sigma^{\rm NP}(z_{\rm cut})=\sigma_{\rm CE}^{\rm NP}(z^*=z_{\rm cut}).
\end{equation}

From Eqs.~(\ref{ace}), (\ref{sceCI}), (\ref{sCI}) and
(\ref{crossdif-pol})--(\ref{F}) some immediate conclusions can be
drawn. First, it is clear that in the case of longitudinally polarized beams
and chosen cut around the beam pipe, $\vert z\vert\le z_{\rm cut}$, the
asymmetry $A_{\rm CE}$ within the SM and in any new physics scenario with
$Z^\prime$ exchanges, and also in the four-fermion contact interaction
scenario, is given by
\begin{equation}
A_{\rm CE}^{\rm SM}=A_{\rm CE}^{\rm SM+CI}
=2 \hskip3pt \frac{z^*({z^*}^2+3)}{z_{\rm cut}(z_{\rm cut}^2+3)}-1.
\label{ACEpol}
\end{equation}  
Secondly, the center--edge asymmetry (\ref{ACEpol}) is identical to that for
unpolarized beams, see Eqs.~(\ref{aceSM}) and (\ref{ACECI}), for $z_{\rm
cut}=1$.  Third, the asymmetry (\ref{ACEpol}) is independent of energy
$\sqrt{s}$, flavour of the final-state fermion $f$, and of the SM and NP
parameters.  Moreover, there is a value $z^*_0$ for which $A_{\rm CE}^{\rm
SM}$ vanishes. One obtains $z^*_0=a-{a}^{-1}$, where
$a=[(p+\sqrt{p^2+4})/2]^{1/3}$, and $p=(3z_{\rm cut}+z_{\rm cut}^3)/2$. These
zeros of $A_{\rm CE}^{\rm SM}$ are important, since the graviton exchange will
there give the only contribution.  Finally, $\sigma^{\rm INT}(z_{\rm cut})=0$
at $z_{\rm cut}=1$, and in this limit for the angular cut the contribution to
the total polarized cross section from the graviton exchange term would be of
order $f_G^2$, i.e., of order $(s/M_H^2)^4$, hence negligible.
\section{Sensitivity}
In order to get some feeling for the sensitivities of the processes 
$e^+e^-\to\mu^+\mu^-$, $b\bar{b}$ and $c\bar{c}$ to graviton exchange effects,
let us consider the statistical significance defined as
\begin{equation}
{\cal S}=\frac{|\Delta A_{\rm CE}|}{\delta A_{\rm CE}},
\label{sens}
\end{equation}
where $\Delta A_{\rm CE}$ is defined by Eq.~(\ref{deviat}).
Here, $\delta A_{\rm CE}$ is the expected statistical uncertainty
defined by Eq.~(\ref{uncert}).  Fig.~\ref{Fig:fig-s-unpol} shows the
statistical significance ${\cal S}$ as a function of $z^*$ for unpolarized
beams for the process (\ref{proc}) at $M_H=2\text{ TeV}$, $\Lumint=50\text{
fb}^{-1}$, $\lambda=1$, and $\sqrt{s}=500\text{ GeV}$.  In the sequel, we
shall put $\lambda=1$; our numerical results will turn out not to depend
appreciably on the choice of the sign.
\begin{figure}[htb]
\refstepcounter{figure}
\label{Fig:fig-s-unpol}
\addtocounter{figure}{-1}
\begin{center}
\setlength{\unitlength}{1cm}
\begin{picture}(10.0,6.0)
\put(0,-0.5)
{\mbox{\epsfysize=7.5cm
\epsffile{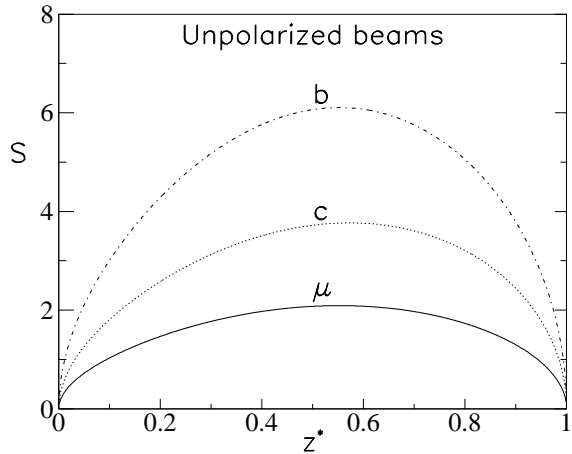}}}
\end{picture}
\caption{Statistical significance, ${\cal S}$, for unpolarized beams,
$M_H=2\text{ TeV}$, $\Lumint=50\text{fb}^{-1}$, $\lambda=1$, and
$\sqrt{s}=500\text{ GeV}$.  Different fermionic final states are considered:
$\mu^+\mu^-$, $c\bar c$ and $b\bar b$. Here, no cut is imposed, $z_{\rm
cut}=1$.}
\end{center}
\end{figure}

From Eqs.~(\ref{sens}), (\ref{delace}), (\ref{uncert}) and
(\ref{aceSM}) one can derive the
statistical significance for unpolarized initial beams
limiting to the interference contribution (and for $z_{\rm cut}=1$):
\begin{equation}  \label{sens_unpol}
{\cal S}_f
=f_G\,{\cal S}_0\,\frac{  
\left|\left({\cal M}_{\rm LL}^{\rm SM}
-{\cal M}_{\rm LR}^{\rm SM}\right)+
\left({\cal M}_{\rm RR}^{\rm SM}-{\cal M}_{\rm RL}^{\rm SM}\right)\right|
}
{\sqrt{ 
\left[({\cal M}_{\rm LL}^{\rm SM})^2
+({\cal M}_{\rm LR}^{\rm SM})^2\right]
+\left[({\cal M}_{\rm RR}^{\rm SM})^2
+({\cal M}_{\rm RL}^{\rm SM})^2\right]
}},
\end{equation}
\begin{equation}
\label{S0}
{\cal S}_0
=\sqrt{\frac{3\pi\alpha_{e.m.}^2\,N_C\,\epsilon_f\Lumint}{s}
\,\frac{z^*(1-z^*)}{(z^*{}^2+3)(z^*{}^2+z^*+4)}
}\, 2(1+z^*).
\end{equation}

The extension of Eq.~(\ref{sens_unpol}) for polarized beams is straightforward:
\begin{equation}
{\cal S}_f
=f_G\,{\cal S}_0\,\sqrt{D}\,\frac{  
\left|(1-P_{\rm eff})\left({\cal M}_{\rm LL}^{\rm SM}
-{\cal M}_{\rm LR}^{\rm SM}\right)+(1+P_{\rm eff})
\left({\cal M}_{\rm RR}^{\rm SM}-{\cal M}_{\rm RL}^{\rm SM}\right)\right|
}
{\sqrt{ 
(1-P_{\rm eff})\left[({\cal M}_{\rm LL}^{\rm SM})^2
+({\cal M}_{\rm LR}^{\rm SM})^2\right]
+(1+P_{\rm eff})\left[({\cal M}_{\rm RR}^{\rm SM})^2
+({\cal M}_{\rm RL}^{\rm SM})^2\right]
}}.
\label{sens_pol}
\end{equation}

Note that the maximum of ${\cal S}$ occurs at $z^*_{\rm max}\approx
1/\sqrt{3}=0.577$ ($\theta\approx 54.7^{\circ}$) which is very close to
$z^*_0$ where $A_{\rm CE}^{\rm SM}=0$.\footnote{Strictly, $1/\sqrt{3}$ 
would be the value of $z^*$ for which 
$\Delta A_{\rm CE}$ in Eq.~(\ref{delace}) is maximal. This represents to a very
good approximation the location $z^*_{\rm max}$ of the maximum of the
statistical significance (\ref{sens}).}
The dependence of ${\cal S}$ in the
vicinity of $z^*_{\rm max}$ is quite smooth as implied by the behaviour of the 
$\Delta A_{\rm CE}$ and $\delta A_{\rm CE}$ shown in
Fig.~\ref{Fig:fig2}. In other words, variation of
$z^*$ around $z^*_{\rm max}$ changes the sensitivity very little. Therefore,
no stringent requirements on angular resolution are needed.

The statistical significance is expressed in terms of the SM amplitudes ${\cal
M}_{\alpha\beta}^{\rm SM} = Q_eQ_f[1+(g_{\alpha}^e g_{\beta}^f/Q_e
Q_f)\,\chi_Z]$. The factor $Q_eQ_f$ is here extracted since it cancels in the
ratios of Eqs.~(\ref{sens_unpol}) and (\ref{sens_pol}).

In order to clarify the dominant role of $q\bar{q}$-pair production over
$\mu^+\mu^-$ production in searching for graviton exchange effects, as shown
in Fig.~\ref{Fig:fig-s-unpol}, and also to reveal the role of polarization in
such analysis, it is instructive to estimate the SM amplitudes in the limit
where
\begin{equation}
s^2_W=0.25, \qquad  M_Z^2\ll s\ll M_H^2.  
\label{approx}
\end{equation}
With these approximations, the relations between the SM couplings 
can be written as
\begin{equation}
 \frac{g_{\rm L}^e}{Q_e}
=\frac{1}{2}\frac{g_{\rm L}^c}{Q_c}
=\frac{1}{5}\frac{g_{\rm L}^b}{Q_b}
=\frac{1}{\sqrt{3}},\qquad
\frac{g_{\rm R}^e}{Q_e}
=\frac{g_{\rm R}^c}{Q_c}
=\frac{g_{\rm R}^b}{Q_b}
=-\frac{1}{\sqrt{3}},
\label{coupl}
\end{equation}
and the SM amplitudes are related as ($\chi_Z\approx 1$)
\begin{align}
  \frac{1}{2}{\cal M}_{\rm LL}^{\rm e\mu}
=&\frac{1}{2}{\cal M}_{\rm RR}^{\rm e\mu}
={\cal M}_{\rm LR}^{\rm e\mu}
={\cal M}_{\rm RL}^{\rm e\mu}
=Q_eQ_\mu\hskip 2pt\frac{2}{3},
\nonumber \\
  \frac{1}{5}{\cal M}_{\rm LL}^{\rm ec}
=&\frac{1}{4}{\cal M}_{\rm RR}^{\rm ec}
= \frac{1}{2}{\cal M}_{\rm LR}^{\rm ec}
={\cal M}_{\rm RL}^{\rm ec}=Q_eQ_c\hskip 2pt\frac{1}{3},
\nonumber \\
  \frac{1}{4}{\cal M}_{\rm LL}^{\rm eb}
=&\frac{1}{2}{\cal M}_{\rm RR}^{\rm eb}
={\cal M}_{\rm LR}^{\rm eb}
=-{\cal M}_{\rm RL}^{\rm eb}
=Q_eQ_b\hskip 2pt\frac{2}{3}.
\label{relat} 
\end{align} 
For unpolarized $e^+e^-$ beams, we have:
\begin{equation}
{\cal S}_\mu:{\cal S}_c:{\cal S}_b
=1:\sqrt{\epsilon_c\frac{135}{23}}:\sqrt{\epsilon_b\frac{135}{11}}
\approx 1:1.9:3.1.
\label{unpol}
\end{equation}
Comparison of the ratios (\ref{unpol}) obtained in the adopted approximation
(\ref{approx}) with those presented in Fig.~\ref{Fig:fig-s-unpol} and derived
from the full expression of Eq.~(\ref{sens}) shows that this approximation is
quite reasonable.  With fully polarized beams, $e^+_{\rm L}e^-_{\rm R}$
($P_{\rm eff}=1$) and $e^+_{\rm R}e^-_{\rm L}$ ($P_{\rm eff}=-1$), we find
\begin{equation}
{\cal S}_\mu(P_{\rm eff}=0):{\cal S}_\mu(P_{\rm eff}=1):
{\cal S}_\mu(P_{\rm eff}=-1)
=1:\sqrt{2}:\sqrt{2}=1:1.4:1.4,
\label{polmm}
\end{equation}
\begin{equation}
{\cal S}_c(P_{\rm eff}=0):{\cal S}_c(P_{\rm eff}=1):{\cal S}_c(P_{\rm eff}=-1)
=1:\sqrt{\frac{46}{17}}:\sqrt{\frac{46}{29}}=1:1.6:1.3,
\label{polcc}
\end{equation}
\begin{equation}
{\cal S}_b(P_{\rm eff}=0):{\cal S}_b(P_{\rm eff}=1):{\cal S}_b(P_{\rm eff}=-1)
=1:\sqrt{\frac{22}{5}}:\sqrt{\frac{22}{17}}=1:2.1:1.1.
\label{polbb}
\end{equation}
Note that the $b\bar{b}$ channel becomes more sensitive to graviton
exchange effects both for unpolarized and polarized beams and would carry
large statistical weight in the analysis.
The advantage of polarization is lessened by the fact that the signal behaves
as $(\sqrt{s}/M_H)^4$ compared to, e.g., the case of four-fermion {\it contact
interactions}. This high power reduces the considerable gain in sensitivity
to a less dramatic 20\% gain in reach on $M_H$ for the $b\bar{b}$ case, see
Eq.~(\ref{polbb}).

The sign of the SM--NP interference term in the $A_{\rm CE}$ asymmetry for the
process $e^+e^-\to c\bar{c}$ is opposite to those of $e^+e^-\to\mu^+\mu^-$ and
$e^+e^-\to b\bar{b}$. This sign correlation might yield additional information
to identify graviton exchange effects.
\section{Identification reach}
To assess a realistic reach on the mass scale $M_H$ we can consider a
$\chi^2$-function made of the deviation of the asymmetry $A_{\rm CE}$ from its
SM value. For a fixed integrated luminosity this can be done using the
statistical errors as well as the systematic errors. We find that, to a very
large extent, the systematic errors associated with the uncertainties expected
on the luminosity measurements cancel out, and the same is true for the
systematic errors induced by the uncertainty on beam polarizations.
Accordingly, the errors on $A_{\rm CE}$ are largely dominated by statistics.
In this estimation we assume the values $\delta\Lumint/\Lumint= \delta
P/P=\delta\bar{P}/\bar{P}=0.5\%$.  We take the beam polarization to be 80\%
and 60\% for electrons and positrons, respectively, and employ a $10^\circ$
angular cut around the beam pipe, i.e., $z_{\rm cut}=0.98$.  Since most of the
error is statistical in origin, we expect the bound on $M_H$ to scale as
$\sim(\Lumint s^3)^{1/8}$.  The dependence of the reach on $M_H$ on $z_{\rm
cut}$ varying in a reasonable range close to 1 is, for the chosen values of
energy, luminosity and polarization, quite smooth.  For example, in the range
$z_{\rm cut} = 0.96-1$, the bound on $M_H$ is found to vary by only a few
percent.

In the present analysis we also take into account the radiative
corrections. Among the complete ${\cal O}(\alpha)$ corrections to the process
(\ref{proc}), the numerically largest QED corrections are the effects of
initial state radiation, which in general are of major importance for new
physics searches. The initial state corrections have been calculated in the
flux function approach (see, {\it e.g.}, ref.~\cite{Hewett:1988xc}). The
structure of the corrected differential cross section in terms of $z_{\rm
c.m.}\equiv\cos\theta$ (where $\theta$ now refers to the final-state
$f\bar{f}$ c.m.\ frame) is \cite{physicsatlep2}
\begin{equation}
\label{Eq:ddsigma}
\frac{\dd\sigma}{\dd z_{\rm c.m.}}
\propto
(1+z_{\rm c.m.}^2)\,\sigma_s +2z_{\rm c.m.}\,\sigma_a,
\end{equation}
The symmetric and antisymmetric parts of the cross section are given by
convolutions of the non-radiative cross section with the flux functions
$H^e_A(v)$, with $v$ the energy of the emitted photon in units of the beam
energy.  Due to the radiative return to the $Z$ resonance for $\sqrt{s}>M_Z$
the energy spectrum of the radiated photons is peaked around $E_\gamma/E_{\rm
beam}\approx 1-M^2_Z/s$.  In order to increase a possible new physics signal,
events with hard photons should be removed by a cut on the photon energy,
$\Delta=E_\gamma/E_{\rm beam}< 1-M^2_Z/s$, with $\Delta=0.9$.  We also take
into account electroweak corrections to the propagators and vertices amounting
essentially to effective (momentum-dependent) coupling constants (effective
Born approximation \cite{Consoli:1989pc} with $m_{\rm top}=175~\text{GeV}$ and
$m_{\rm higgs}=300~\text{GeV}$).  Concerning the other QED corrections, the
final state ones and the initial-final state interference, they can be checked
to be numerically unimportant for the chosen kinematical cuts, in particular
that on $\Delta$, using the existing codes, e.g., ZFITTER
\cite{Bardin:2001yd}. In addition $z^*_0$, the zero of $A_{\rm CE}^{\rm SM}$,
is shifted by these corrections by a little amount from the effective Born
approximation value. The box-diagram contributions, which introduce a
different angular dependence, are found to be very small.

Since the form of the corrected cross section, Eq.~(\ref{Eq:ddsigma}), is the
same as that of Eq.~(\ref{crossdif}), it follows that the
radiatively-corrected zero of $A_{\rm CE}^{\rm SM}$, $z^*_0$, can again be
defined by:
\begin{equation}
\left[\int_{-z^*_0}^{z^*_0}
-\left(\int_{-1}^{-z^*_0}+\int_{z^*_0}^{1}\right)\right]
(1+z^2){\dd z}=0, 
\label{eq_z0} 
\end{equation} 
and one finds the same value for $z^*_0$ as given by
Eq.~(\ref{z*0}). Moreover, in both the SM and SM+CI cases the radiatively
corrected asymmetry $A_{\rm CE}$ is still determined by Eq.~(\ref{ACECI}).

Summing over $\mu^+\mu^-$, $\tau^+\tau^-$, $b\bar b$ and $c\bar c$ final
states (the top quark is excluded as its mass effects would alter the angular
distribution (\ref{Eq:ddsigma})) one can perform a conventional $\chi^2$
analysis:
\begin{equation}
\chi^2=\sum_{f=\mu, \tau, c, b}
\frac{(\Delta A_{\rm CE}^f)^2}{(\delta A_{\rm CE}^f)^2},
\end{equation}
keeping $z^*=z^*_0$ fixed (recall from Fig.~\ref{Fig:fig-s-unpol} that the
sensitivities for the various final states are rather smooth in an interval
around $z^*_0\simeq z^*_{\rm max}$).  This leads to the $5\sigma$
identification reach as a function of integrated luminosity with energy
$\sqrt{s}=0.5$, 1, 3 and 5~TeV shown in Fig.~\ref{Fig:fig4}. The chosen range
of energy corresponds to TESLA, NLC \cite{Aguilar-Saavedra:2001rg} and CLIC
\cite{Assmann:2000hg}.  Specifically, for $\sqrt{s}=0.5-1$~TeV and $3-5$~TeV
machines with integrated luminosity $1~\text{ab}^{-1}$ the identification
reach with double beam polarization is found to be $(7-6)\times\sqrt{s}$ and
$(4.5-4)\times\sqrt{s}$, respectively.  The effects of spin-2 graviton
exchange can be distinguished from the other forms of contact-interaction-like
effects considered in Table~\ref{table:epsilon} for $M_H\le 3.5,\ 6,\
13.6$ and 20~TeV at $\sqrt{s}=0.5$, 1, 3 and 5~TeV, respectively.
\begin{figure}[htb]
\refstepcounter{figure}
\label{Fig:fig4}
\addtocounter{figure}{-1}
\begin{center}
\setlength{\unitlength}{1cm}
\begin{picture}(10.0,6.0)
\put(-4,-0.5)
{\mbox{\epsfysize=7.5cm\epsffile{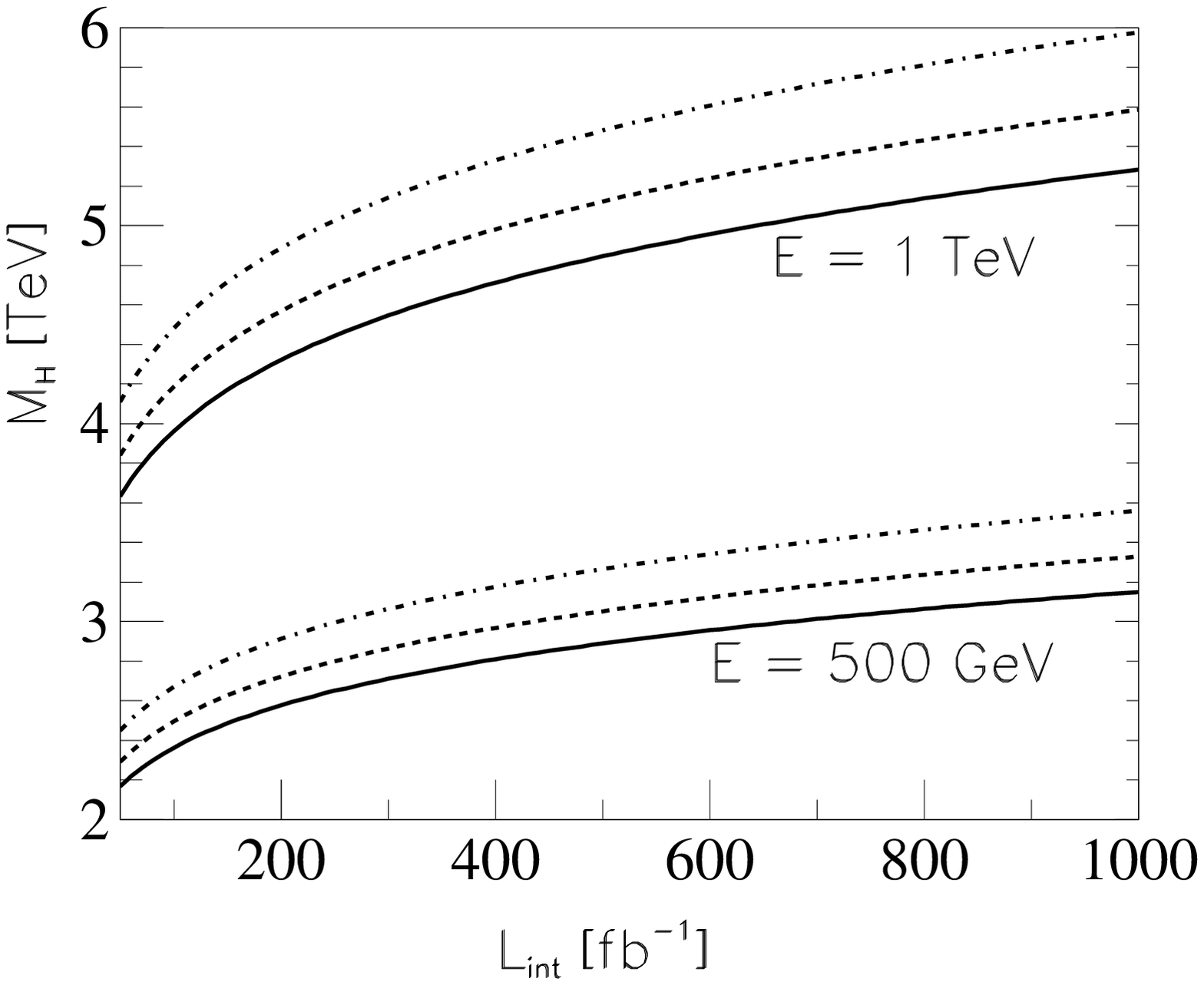}}
 \mbox{\epsfysize=7.5cm\epsffile{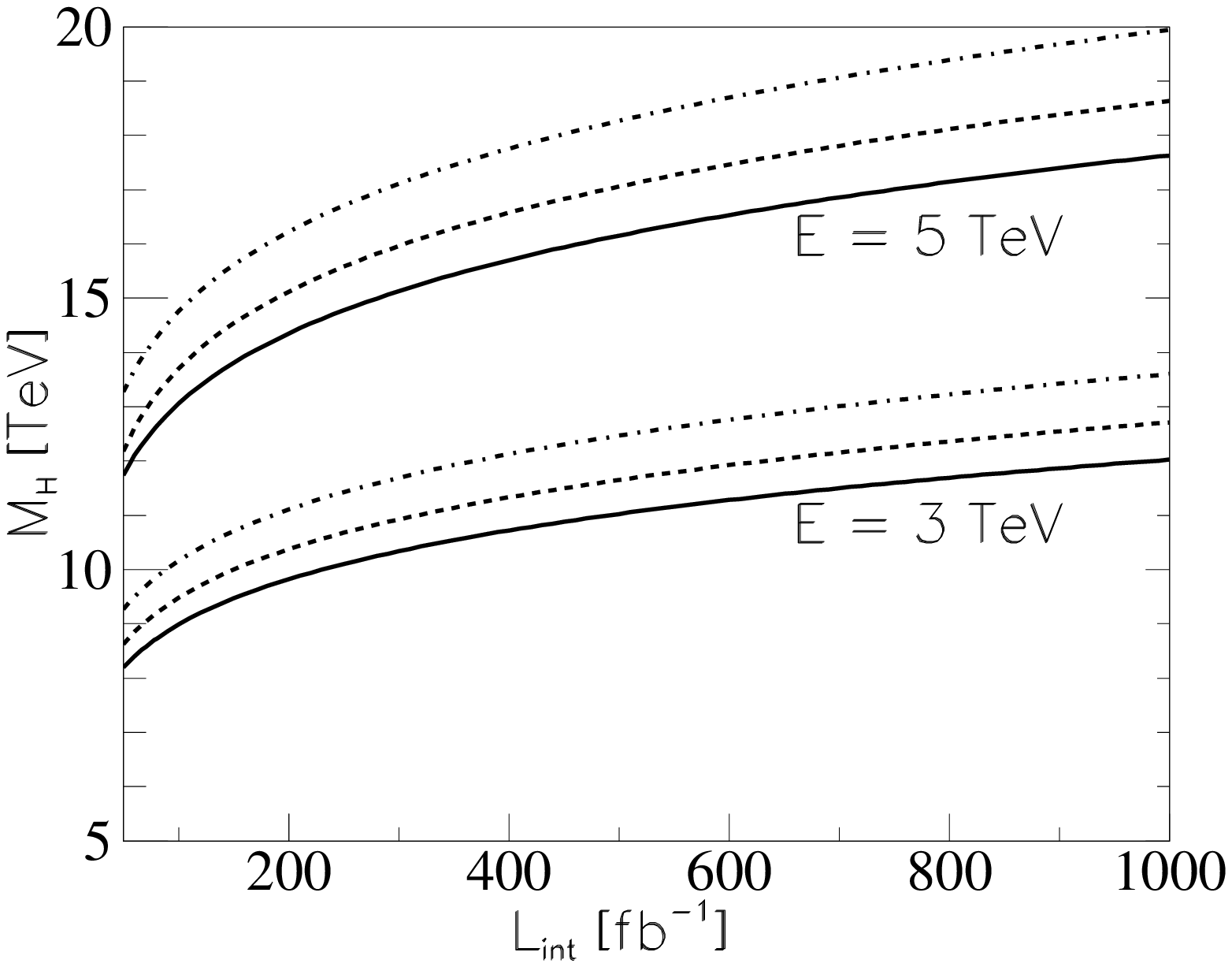}}}
\end{picture}
\caption{5$\sigma$ reach on the mass scale $M_H$ {\it vs.}\ integrated
luminosity from the process $e^+e^-\to f{\bar f}$, with $f$ summed over
$\mu,\tau,b,c$, and for a range of energies from 0.5 to 5~TeV.
Solid: unpolarized; dashed: electrons polarized, $P=0.8$;
dash-dotted: both beams polarized, $P=0.8$, $\bar P=-0.6$.}
\end{center}
\end{figure}

It turns out that under the assumption of no observation of $\Delta A_{\rm
CE}$ within the expected experimental uncertainty, in which case only bounds
on $f_G$ can be derived, the 95\% CL lower limits on $M_{\rm H}$ would be
represented by the values shown in Fig.~\ref{Fig:fig4} essentially multiplied
by a factor of the order of 1.3.

Finally, we consider a scenario that would most closely mimic massive graviton
exchange, namely the exchange of a scalar field in the $s$- and $t$-channels,
limiting ourselves to the production of lepton pairs. To be specific, we
can concentrate on the example $R$-parity breaking SUSY interactions 
mediated by sneutrino
exchange \cite{Kalinowski:1997bc,Rizzo:1998vf}.  First, we consider the
$t$-channel $\tilde\nu$ contribution to $e^+e^-\to\mu^+\mu^-$ or
$\tau^+\tau^-$. In this case the helicity cross sections are given by
Eq.~(\ref{helcross}) with an additional contribution to the helicity
amplitudes caused by $\tilde\nu$ exchange:
\begin{equation}
\Delta_{\rm LL}=\Delta_{\rm RR}=0, \qquad \Delta_{\rm LR}=\Delta_{\rm RL}
=\frac{1}{2}C_{\tilde\nu}P^t_{\tilde\nu},
\label{delta_nu}
\end{equation} 
where $P^t_{\tilde\nu}= s/(t- m_{\tilde\nu}^2)$ and $t=-s(1-z)/2$,
$C_{\tilde\nu}=\lambda^2/4\pi\alpha_{\rm e.m.}$, with $\lambda$ in this case
the Yukawa coupling \cite{Rizzo:1998vf}. It is clear that in the contact
interaction limit, i.e.  $\vert t\vert\ll m_{\tilde\nu}^2$, these two new
physics effects, graviton exchange and $\tilde\nu$ exchange, are easily
separable by the previous analysis based on the asymmetry $A_{\rm CE}$.  If we
are not in the contact interaction limit, ${\cal M}_{\rm LR}$ and ${\cal
M}_{\rm RL}$ pick up an additional $z$ dependence resulting in a $z^*$
dependence of $\Delta A_{\rm CE}$ different from the one in Eq.~(\ref{delace})
determined by graviton exchange.  We find that polarization will also help to
distinguish these two new physics effects. For this purpose one can define the
polarized observable, the absolute center-edge left-right asymmetry:
\begin{equation}
\sigma A_{\rm CE,LR}\equiv
\sigma_{\rm CE,LR}=\left[\int_{-z^*_0}^{z^*_0}-
\left(\int_{-1}^{-z^*_0}+\int_{z^*_0}^{1}\right)\right]
\left( \frac{\dd\sigma_{\rm L}}{\dd z}
-\frac{\dd\sigma_{\rm R}}{\dd z}\right)\,{\dd z}.
\label{sceLR} 
\end{equation} 
Here, 
$z^*_0$ is the zero of 
$A_{\rm CE}^{\rm SM}$, see Eq.~(\ref{eq_z0}), and
$\dd\sigma_{\rm L}/{\dd z}$ and $\dd\sigma_{\rm R}/{\dd z}$ are the
differential cross sections defined by Eq.~(\ref{crossdif-pol}) with specific
choices of electron and positron beam polarizations, for example $(P,\bar
P)=(-P_1,P_2)$ and $(P_1,-P_2)$, respectively, with $P_1$ and $P_2$
positive. The deviation from the SM prediction of the differential cross
section difference involved in Eq.~(\ref{sceLR}) and caused by $\tilde\nu$
exchange is given by
\begin{equation}
\Delta\frac{\dd\sigma_{\rm LR}}{\dd z}\equiv
\left(\frac{\dd\sigma_{\rm L}}{\dd z}-\frac{\dd\sigma_{\rm R}}{\dd z}\right)
-\left(\frac{\dd\sigma_{\rm L}^{\rm SM}}{\dd z}
-\frac{\dd\sigma_{\rm R}^{\rm SM}}{\dd z}\right)
\propto P_{\rm eff}\,({\cal M}_{\rm LR}^{\rm SM}-{\cal M}_{\rm RL}^{\rm SM})
C_{\tilde\nu}P^t_{\tilde\nu}=0,
\label{difer_LR} 
\end{equation} 
because ${\cal M}_{\rm LR}^{\rm SM}={\cal M}_{\rm RL}^{\rm SM}$ for the
process (\ref{proc}) with $f=\mu,\tau$.  Notice that this property, easily
checked in the tree approximation of the SM, continues to hold also in the
effective Born approximation. Accordingly, $\sigma_{\rm CE,LR}$ is unaltered
by sneutrino exchange in the leptonic processes $e^+e^-\to\mu^+\mu^-$ and
$e^+e^-\to\tau^+\tau^-$, i.e., $\Delta\sigma_{\rm CE,LR}^{\tilde\nu}=0$,
whereas it is modified by graviton exchange, $\Delta\sigma_{\rm CE,LR}^{G}\ne
0$.  The choice of $z^*_0$ as integration limits in (\ref{sceLR}) assures that
the contribution of the SM as well as that of any conventional contact
interaction vanish, leaving room only for graviton and sneutrino exchanges.
The role of polarization is that, in the combination (\ref{sceLR}), the
sneutrino contributions cancel as explicitly seen in (\ref{difer_LR}), so that
only the signal of the graviton exchange term can survive. One can notice that
this kind of analysis is allowed also in the case of only electron beam
longitudinal polarization and unpolarized positron, namely, $P_1\ne 0$ and
$P_2=0$. Also, the quadratic term in the differential cross sections,
proportional to $(C_{\tilde\nu}P^t_{\tilde\nu})^2$, cancels in
Eq.~(\ref{sceLR}), so that Eq.~(\ref{difer_LR}), linear in
$(C_{\tilde\nu}P^t_{\tilde\nu})$ is the exact representation of the deviation
from the SM.

Concerning ${\tilde\nu}$ exchange in the $s$ channel, the polarized
differential cross section (\ref{crossdif-pol}) picks up an additional,
$z$-independent, term:
\begin{equation}
\frac{\dd\sigma^s}{\dd z}
\propto(1+P\bar{P})(C_{\tilde\nu}P^s_{\tilde\nu})^2,
\label{nu_s}
\end{equation}
with $P^s_{\tilde\nu}\simeq s/(s- m_{\tilde\nu}^2)$.  Indeed, the $s$-channel
scalar exchange diagram does not interfere with the electroweak SM amplitudes
mediated by the $\gamma$ and $Z$ boson and the resulting effects are of
quadratic order, $(C_{\tilde\nu}P^s_{\tilde\nu})^2$. As is easily seen from
Eq.~(\ref{nu_s}), either the electron beam polarization or both electron and
positron polarizations allow to remove the sneutrino $s$ channel exchange
contribution to Eq.~(\ref{sceLR}), i.e., $\Delta\sigma_{\rm
CE,LR}^{\tilde\nu}=0$ also in this case.

Conversely, it is possible to define an observable `orthogonal' to
$\sigma_{\rm CE,LR}$ which is sensitive to $\tilde\nu$ exchange in the $s$
channel and independent of the effects of {\it graviton exchange}, {\it
contact interactions}, and $Z'$ exchange. This is the double beam polarization
asymmetry defined as \cite{Rizzo:1998vf}
\begin{equation}
A_{\rm double}=\frac
{\sigma(P_1,-P_2)+\sigma(-P_1,P_2)-\sigma(P_1,P_2)-\sigma(-P_1,-P_2)}
{\sigma(P_1,-P_2)+\sigma(-P_1,P_2)+\sigma(P_1,P_2)+\sigma(-P_1,-P_2)}.
\label{double}
\end{equation}
Here, $\sigma$ are the cross sections integrated over $z$ in the indicated
polarization configurations. One can see immediately that for the case of the
SM, {\it contact interactions}, $Z'$ exchange, $\tilde\nu$ exchange in the $t$
channel and for {\it graviton exchange} one obtains $A_{\rm double}=P_1P_2$
since these exchanges contribute to the same amplitudes, whereas $\tilde\nu$
exchange in the $s$ channel will force this observable to smaller values as
$\Delta A_{\rm double}\propto -P_1P_2\,(C_{\tilde\nu}P^s_{\tilde\nu})^2<0$.
A value of $A_{\rm double}$  smaller than $P_1P_2$ can provide
a signature of scalar exchange in the $s$ channel.

In conclusion, we have seen that one can define a set of observables using
cross sections integrated within appropriate angular limits that can
discriminate among deviations from the SM prediction related either to
graviton or to scalar exchange in the $s$ channel.

\section{Summary and observations}
We conclude with a summary of the main points and some observations.
We have developed a specific approach based on an integrated observable,
the center-edge asymmetry $A_{\rm CE}$, to search for and identify spin-2 graviton 
exchange with uniquely distinct signature. Indeed, 
the spin-2 graviton KK exchanges contribute to the asymmetry
$A_{\rm CE}$, whereas no deviation from the SM
is induced by other kinds of new 
physics such as the composite-like contact interactions, 
a heavy vector boson $Z^\prime$, gauge boson KK excitations listed in Table~1.
Both in the SM and in any new physics scenario described by effective
current--current interactions, the asymmetry $A_{\rm CE}$  is identical
for any value of the parameter $z^*$.

Particularly convenient is the range of $z^*$ values around the zero of $A_{\rm
CE}$ ($z^*_0$) for the SM. In this range, the sensitivity of $A_{\rm CE}$ to
the graviton coupling $f_G$ is maximal and rather smooth in $z^*$, so that one
can obtain not only the discovery but the real unambiguous identification of
this new physics effect. This kind of analysis based on $A_{\rm CE}$ can be
applied also to the case where a cut is imposed on the full angular range
covered by the experiment, and its nice distinctive features continue to hold
to a very good approximation.

Initial electron and positron beam polarization appears to increase the
sensitivity to graviton exchange, but their impact on the mass scale parameter
$M_{\rm H}$ is not dramatic due to the large power $(\sqrt{s}/M_{\rm H})^4$
that parametrizes the graviton coupling.  In particular, for an $e^+e^-$
linear collider with energy $\sqrt{s}=0.5$, 1, 3 and 5~TeV, with integrated
luminosity $1~\text{ab}^{-1}$, double beam polarization and a $10^\circ$
angular cut, the $5\sigma$ identification reach is found to be $M_H\le 3.5,\
6,\ 13.6$ and 20~TeV, respectively.

Instead, initial polarization can play a key role in distinguishing graviton
exchange from competing effects, such as those originating from exchange of
scalar particles, for which appropriate polarization asymmetries can be
defined.

An approach aiming to isolate graviton-exchange effects has recently been
proposed in Ref.~\cite{Rizzo:2002pc}, based on the differential cross section
convoluted with Legendre polynomials and integrated over the angular range.
Alternatively, our method directly uses the integrated cross sections to
construct the center-edge asymmetry $A_{\rm CE}$.  It has the main advantage
of a mild dependence of $A_{\rm CE}$ on the kinematical cut, systematics, and
on the number of angular bins, and in particular it depends on the total
luminosity and not on the statistics available in each bin. These features may
lead to some improvement in the 5-$\sigma$ discovery reach on the mass scale
$M_H$.

Finally, we note that an analysis based on asymmetries analogous to 
$A_{\rm CE}$  might be useful in the context of hadronic collisions, 
in the Drell--Yan process.
\goodbreak

\medskip
\leftline{\bf Acknowledgement}
\par\noindent
This research has been partially supported by MIUR (Italian Ministry of
University and Research), by funds of the University of Trieste,
and by the Research Council of Norway.



\begin{thebibliography}{99}
\bibitem{'tHooft:xb}
G.~'t Hooft, in ``Recent Developments In Gauge Theories'', Proceedings, 
Nato Advanced Study Institute, Cargese, France, August 26 - September 8, 1979,
edited by
G.~'t Hooft, C.~Itzykson, A.~Jaffe, H.~Lehmann, P.~K.~Mitter, 
I.~M.~Singer and R.~Stora
{\it  New York, USA: Plenum (1980) 
(Nato Advanced Study Institutes Series: Series B, Physics, 59)}; \\
S.~Dimopoulos, S.~Raby and L.~Susskind,
Nucl.\ Phys.\ B {\bf 173} (1980) 208.

\bibitem{Eichten:1983hw}
E.~Eichten, K.~D.~Lane and M.~E.~Peskin,
Phys.\ Rev.\ Lett.\  {\bf 50} (1983) 811; \\
R.~Ruckl,
Phys.\ Lett.\ B {\bf 129} (1983) 363.

\bibitem{Barger:1997nf}
V.~D.~Barger, K.~m.~Cheung, K.~Hagiwara and D.~Zeppenfeld,
Phys.\ Rev.\ D {\bf 57} (1998) 391
[arXiv:hep-ph/9707412]; \\
D.~Zeppenfeld and K.~m.~Cheung,
Proceedings of 5th International WEIN Symposium: 
A Conference on Physics Beyond the Standard Model (WEIN 98), Santa Fe, NM, 
14--21 Jun 1998;
preprint MADPH-98-1081
[arXiv:hep-ph/9810277].

\bibitem{Hewett:1988xc}
For reviews see, {\it e.g.}:
J.~L.~Hewett and T.~G.~Rizzo,
Phys.\ Rept.\  {\bf 183} (1989) 193; \\ 
A.~Leike,
Phys.\ Rept.\  {\bf 317} (1999) 143
[arXiv:hep-ph/9805494];

\bibitem{Buchmuller:1986zs}
W.~Buchmuller, R.~Ruckl and D.~Wyler,
Phys.\ Lett.\ B {\bf 191} (1987) 442
[Erratum-ibid.\ B {\bf 448} (1999) 320];\\
W.~Buchmuller, R.~R\"uckl and D.~Wyler, Phys. Lett. B {\bf 191} 442, 
B {\bf 448} (1999) 320; \\
G.~Altarelli, J.~R.~Ellis, G.~F.~Giudice, S.~Lola and M.~L.~Mangano,
Nucl.\ Phys.\ B {\bf 506} (1997) 3
[arXiv:hep-ph/9703276]; \\
R.~Casalbuoni, S.~De Curtis, D.~Dominici and R.~Gatto, 
Phys.\ Lett.\ B {\bf 460} (1999) 135
[arXiv:hep-ph/9905568];\\
V.~D.~Barger and K.~m.~Cheung,
Phys.\ Lett.\ B {\bf 480} (2000) 149
[arXiv:hep-ph/0002259]. 

\bibitem{Kalinowski:1997bc}
J.~Kalinowski, R.~Ruckl, H.~Spiesberger and P.~M.~Zerwas,
Phys.\ Lett.\ B {\bf 406} (1997) 314
[arXiv:hep-ph/9703436];\\
Phys.\ Lett.\ B {\bf 414} (1997) 297
[arXiv:hep-ph/9708272].

\bibitem{Rizzo:1998vf}
T.~G.~Rizzo,
Phys.\ Rev.\ D {\bf 59} (1999) 113004
[arXiv:hep-ph/9811440].

\bibitem{Cuypers:1996ia}
For a review, see 
F.~Cuypers and S.~Davidson,
Eur.\ Phys.\ J.\ C {\bf 2} (1998) 503
[arXiv:hep-ph/9609487], and references there.

\bibitem{Gounaris:1997ft}
G.~J.~Gounaris, D.~T.~Papadamou and F.~M.~Renard,
Phys.\ Rev.\ D {\bf 56} (1997) 3970
[arXiv:hep-ph/9703281].

\bibitem{Arkani-Hamed:1998rs}
N.~Arkani-Hamed, S.~Dimopoulos and G.~R.~Dvali,
Phys.\ Lett.\ B {\bf 429} (1998) 263
[arXiv:hep-ph/9803315];
Phys.\ Rev.\ D {\bf 59} (1999) 086004
[arXiv:hep-ph/9807344]; \\
L.~Randall and R.~Sundrum,
Phys.\ Rev.\ Lett.\  {\bf 83} (1999) 3370
[arXiv:hep-ph/9905221];\\
I.~Antoniadis, N.~Arkani-Hamed, S.~Dimopoulos and G.~R.~Dvali,
Phys.\ Lett.\ B {\bf 436} (1998) 257
[arXiv:hep-ph/9804398].

\bibitem{Giudice:1998ck}
G.~F.~Giudice, R.~Rattazzi and J.~D.~Wells,
Nucl.\ Phys.\ B {\bf 544} (1999) 3
[arXiv:hep-ph/9811291];
Nucl.\ Phys.\ B {\bf 630} (2002) 293
[arXiv:hep-ph/0112161].

\bibitem{Hewett:1998sn}
J.~L.~Hewett,
Phys.\ Rev.\ Lett.\  {\bf 82} (1999) 4765
[arXiv:hep-ph/9811356];\\
T.~Han, J.~D.~Lykken and R.~J.~Zhang,
Phys.\ Rev.\ D {\bf 59} (1999) 105006
[arXiv:hep-ph/9811350];\\
T.~G.~Rizzo,
Phys.\ Rev.\ D {\bf 64} (2001) 095010
[arXiv:hep-ph/0106336];\\
H.~Davoudiasl, J.~L.~Hewett and T.~G.~Rizzo,
Phys.\ Rev.\ Lett.\  {\bf 84} (2000) 2080
[arXiv:hep-ph/9909255];
Phys.\ Rev.\ D {\bf 63} (2001) 075004
[arXiv:hep-ph/0006041];\\
E.~A.~Mirabelli, M.~Perelstein and M.~E.~Peskin,
Phys.\ Rev.\ Lett.\  {\bf 82} (1999) 2236
[arXiv:hep-ph/9811337];\\
S.~Cullen, M.~Perelstein and M.~E.~Peskin,
Phys.\ Rev.\ D {\bf 62} (2000) 055012
[arXiv:hep-ph/0001166].

\bibitem{Antoniadis:1993jp}
I.~Antoniadis and K.~Benakli,
Phys.\ Lett.\ B {\bf 326}, 69 (1994)
[arXiv:hep-th/9310151];
I.~Antoniadis, K.~Benakli and M.~Quiros,
Phys.\ Lett.\ B {\bf 331}, 313 (1994)
[arXiv:hep-ph/9403290].

\bibitem{Pasztor:2001hc}
G.~Pasztor and M.~Perelstein,
in {\it Proc. of the APS/DPF/DPB Summer Study on the Future of Particle 
Physics (Snowmass 2001) } ed. N.~Graf,
arXiv:hep-ph/0111471.

\bibitem{Cheung:2001mq}
K.~m.~Cheung and G.~Landsberg,
Phys.\ Rev.\ D {\bf 65} (2002) 076003
[arXiv:hep-ph/0110346].

\bibitem{Aguilar-Saavedra:2001rg}
J.~A.~Aguilar-Saavedra {\it et al.}  [ECFA/DESY LC Physics Working Group
                  Collaboration],
``TESLA Technical Design Report Part III: 
Physics at an e+e- Linear Collider,''
DESY-01-011,
arXiv:hep-ph/0106315;\\
T.~Abe {\it et al.}  [American Linear Collider Working Group Collaboration],
``Linear collider physics resource book for Snowmass 2001. 1:  Introduction,''
in {\it Proc. of the APS/DPF/DPB Summer Study 
on the Future of Particle Physics (Snowmass 2001) } ed. N.~Graf,
SLAC-R-570, 
arXiv:hep-ex/0106055.

\bibitem{Assmann:2000hg}
R.~W.~Assmann {\it et al.},
``A 3-TeV e+ e- linear collider based on CLIC technology,''
SLAC-REPRINT-2000-096. 

\bibitem{Rizzo:2002pc}
T.~G.~Rizzo,
JHEP {\bf 0210} (2002) 013
[arXiv:hep-ph/0208027]; \\
T.~G.~Rizzo,
arXiv:hep-ph/0211374.

\bibitem{Schrempp:1987zy}
B.~Schrempp, F.~Schrempp, N.~Wermes and D.~Zeppenfeld,
Nucl.\ Phys.\ B {\bf 296} (1988) 1.

\bibitem{Pankov:1997da}
A.~A.~Pankov, N.~Paver and C.~Verzegnassi,
Int.\ J.\ Mod.\ Phys.\ A {\bf 13} (1998) 1629
[arXiv:hep-ph/9701359].

\bibitem{Gounaris:1992kp}
G.~Gounaris, J.~Layssac, G.~Moultaka and F.~M.~Renard,
Int.\ J.\ Mod.\ Phys.\ A {\bf 8} (1993) 3285.

\bibitem{Datta:2002tk}
A.~Datta, E.~Gabrielli and B.~Mele,
Phys.\ Lett.\ B {\bf 552} (2003) 237
[arXiv:hep-ph/0210318].

\bibitem{Flottmann:1995ga}
K.~Flottmann,
DESY-95-064; 
%
K.~Fujii and T.~Omori,
KEK-PREPRINT-95-127.

\bibitem{physicsatlep2}
For a review see, e.g., W. Beenakker, F. A. Berends (conv.): Proc.
of the Workshop ``Physics at LEP2'', CERN 96-01, vol. 1, p. 79 and
references therein.

\bibitem{Consoli:1989pc}
M.~Consoli, W.~Hollik and F.~Jegerlehner,
CERN-TH-5527-89, presented at the {\it Workshop on Z Physics at LEP}; \\
G.~Altarelli, R.~Casalbuoni, D.~Dominici, F.~Feruglio and R.~Gatto,
Nucl.\ Phys.\ B {\bf 342} (1990) 15.

\bibitem{Bardin:2001yd}
D.~Bardin, P.~Christova, M.~Jack, L.~Kalinovskaya, A.~Olchevski, 
S.~Riemann and T.~Riemann,
Comput.\ Phys.\ Commun.\  {\bf 133} (2001) 229
[hep-ph/9908433].

\end{thebibliography}
\end{document}